# Surface charges effects on the 2D conformation of supercoiled DNA

**Tatiana Schmatko[*], Pierre Muller, and Mounir Maaloum**
Institut Charles Sadron et Université de Strasbourg, CNRS UPR 22,
23 rue du loess, BP 84047, 67034 Strasbourg Cedex2, France

**ABSTRACT**
We have adsorbed plasmid pUc19 DNA on a supported bilayer. The mobility of the lipids within the bilayer ensured a 2D equilibrium of the DNA molecule. By varying the fraction of cationic lipids in the membrane, we have tuned the surface charge. Plasmids conformations were imaged by Atomic Force Microscopy (AFM).We performed two sets of experiments: deposition from salt free solution on charged bilayers and deposition from salty solutions on neutral bilayers. Plasmids can be seen as rings, completely opened structures, or tightly supercoiled plectonemes, depending on the experimental conditions. The plectonemic conformation is observed either on a highly charged surface in the absence of salt or on a neutral bilayer at 30 mM of salt. We demonstrate the equivalence of surface screening by mobile interfacial charges and bulk screening from salt ions.

**INTRODUCTION**
Since its discovery (1), supercoiled DNA has been extensively studied by biologists and by physicists. The former were mostly interested in its implication in biological processes (2-8), the later were amazed by the beauty of its structure and studied its topology and conformation (9,10). Topology and conformation are inter-dependent so that, in order to study the conformation of a supercoiled DNA we need to define a few topological concepts.

A supercoiled DNA is a circular double stranded DNA (ds-DNA) molecule that is constrained by a fixed linking number $L_k$. $L_k$ is a topological invariant. Two other numbers are of great importance: the Twist number $T_W$ and the Writhing number $W_r$. These three quantities are linked by the following equation: $L_k = T_W + W_r$. Because of the natural twist of the DNA double helix, there is an intrinsic linking number $L_{k0}$. Therefore, in practice, linking numbers are always defined with respect to $L_{k0}$ via the linking number difference $\Delta L_k$. As $L_{k0}$ increases with the DNA length, one usually introduces an intensive parameter $\sigma$, the supercoiled density, that enables the comparison of plasmids of different lengths : $\sigma = \dfrac{\Delta L_k}{L_{k0}}$.

For a fixed linking number, twist can be traded against writhe. A supercoiled DNA molecule attempts to relax its torsional stress by writhing. Segments of the chain go successively above and below others which finally defines the tertiary structure of the molecule. Writhe gives rise to extra bending. The conformation is fixed by a balance between twist and bending energies. There is an additional electrostatic energy contribution. The global balance depends on the salt concentration in solution.

Much is known about the conformation of supercoiled DNA in solution, mostly under the physiological conditions relevant in biology. Many experiments using gradient centrifugation (11) and electrophoresis have been performed in the late 70's. They tend to give precise values for $W_r$, $T_W$ and $\sigma$, as well as ratios between them (10,12,13).

The first electron microscopy (EM) images of plasmids were available roughly ten years later (10,14-17). This was the first time one could see what supercoiled DNA looks like. In these experiments, supercoiled DNA is shown to have a plectonemic conformation. For very long DNAs, these plectonemes may have branches as well. It is described as a superhelix of radius $r$. The number of crossings of one strand above or below the other is $n$. $W_r = n \sin \alpha$, where $\alpha$ is the angle between the tangent to one helix and the plane perpendicular to the superhelix axis. One of the works of reference is by Boles et al (10). It shows perfectly regular plectonemes with a particularly well defined superhelix radius. In this study the authors measured $n$, $r$ and $W_r$ precisely and compared their results with those obtained from solution measurements. These observations were confirmed by molecular simulations (18-24) so that all experiments performed later refer to this work.

---

[*] To whom correspondence should be addressed

AFM data arrived later (25-35). They systematically show loosely interwound conformations. The observed conformations are much less regular than those seen in EM. Small variations were observed from experiments to experiments, and sometimes the results were discarded because they did not show plectonemic conformations.

In 2002 Zhakharova et al (36) performed small angle neutron scattering on supercoiled DNA. Their results are in agreement with pioneering EM work. For instance, they obtained in 50 mM NaCl an opening angle α of 50° and a superhelix radius $r$ of 10nm. Surprisingly the superhelix radius distribution was rather large (± 4 nm) in comparison with EM measurements. This cannot be attributed to the precision of the measurement as the resolution of neutron scattering is much better than 4 nm. These results could then go in favor of the AFM loose conformations as well. Neutron or XR scattering do not perturb the molecule and can be trustfully considered, but it is rather difficult to transpose the results to real space. On the other hand, real space experiments like AFM and TEM require the deposition of the molecule at the surface, whose effect on the conformation is often underestimated. People are aware of the influence of the surface, but it is often neglected. In fact it is an important issue. One might think that the problem of surface effects was solved with the appearance of cryo-EM which is surely, a real improvement in this direction (7,37). With cryo-EM, the sample solution is frozen and micro sliced before imaging, such as we may think that the observed conformation is the real 3D conformation. The results obtained with cryo-EM were by many aspects comparable to those from pioneering TEM work. For example the idea of a perfectly regular 3D, plectonemic conformation was reinforced.

Nonetheless, cryo-EM can bring surface artifacts as well. Sample preparation is very subtle and small volumes are required. Because the sample thickness is often of the same order of magnitude as the radius of gyration of the molecule, confinement may occur before freezing at the air/ water interface (37). At large ionic strength, cryo-EM also showed strong deviations from solution measurements; unexpectedly, the gyration radius of the molecule was seen to increase with the salt concentration (14) (17) .

This discrepancy was attributed to a quenching of the conformation at very low temperature. Indeed, the native twist of the DNA molecule should increase when the temperature is decreased. Simulations at – 50°C have confirmed this assumption (38).

Our goal in this article is to understand the influence of surface charges when supercoiled DNA is confined in 2D, and to help rationalize seemingly conflicting observations reported in the literature. The physics of the problem is complex. There is entropy loss consecutive to DNA adsorption, entropy gain consecutive to counter-ions release, and various electrostatic interactions. For constrained cyclic DNA, supercoiling introduces an extra level of complexity. All these aspects enter the description with different importance depending on the quantity of salt in solution. Vologodski and Cozzarelli (39) performed simulations in order to elucidate the influence of the surface on the conformation of supercoiled DNA. This work only considered non equilibrium immobilization. Later on, Velichko and coworkers (40) did molecular simulations on supercoiled semi flexible polymers and reported writhe gain when the macromolecule was adsorbed on a surface. The model used to describe the supercoiled macromolecule was a freely-joined chain with non-charged segments. Their study thus only took into account the entropic aspect and their conclusions are not directly applicable to DNA.

In a more recent Monte Carlo simulation, Fujimoto and Schurr (41) used a model previously validated with supercoiled DNA in solution. They added a surface potential to mimic the 2D immobilization of the molecule. Our own experiments will turn out to be quite close to their simulation results at equilibrium. The authors compared their computation with the AFM results from Lyubchenko and Shlyakhtenko (29) even though neither the surface charge of the amino- modified mica, nor the plasmid supercoiled density, were known. Their results did not perfectly match the experiments. They do coincide with the AFM data at 161 mM salt and show that after surface immobilization, the DNA molecule seems to rearrange in a linear fashion. At 10 mM however, they show plectonemic conformations, while experiments were reporting quite open conformations, with multiple strands crossing nearly at the same point. The authors assumed that the unusual multi-arm conformations reported in AFM might be due to irregularities of the surface used for DNA fixing, or local higher charge densities.

Starting from this statement, Bussiek and Langowsky (33) performed AFM experiments where the DNA was deposited on polylysine films. The polymer density was varied. Testing linear DNA with the model from Rivetti et al (42) on their surfaces, they showed that for high polylysine densities, DNA was immediately stuck and remained in a conformation that is a projection of the full 3D conformation. For plasmids of the same length they observed conformations that were irregular and frozen. For low polylysine densities, the linear DNAs reached an equilibrated 2D conformation and plasmids were observed with plectonemic conformations. They concluded that the plectonemes were the result of a 2D equilibrium conformation and that open or multi-arm or many crossing conformations were the result of a 3D freezing.

This work is a good starting point toward the understanding of surface induced conformations, although electrostatics is not taken into account. Surface charges were only considered as an adhesion promoter.

In the present article we present AFM experiments performed on surfaces with known surface charge densities. By adsorbing DNA on lipid bilayers whose composition was a mixture of cationic lipids and zwitterionic phospholipids, we have varied the surface charge quantitatively. Furthermore, lipids are mobile within a membrane which ensures that DNA molecules are not frozen and conformations should be equilibrated in 2D. The lipids diffusion coefficient has been measured by Fluorescence Recovery After Patterned Photobleaching FRAPP (43). Our membranes are in the gel phase where the dynamics are 1000 times slower than in the fluid phase, but still fast enough to allow small segments of the DNA chain to move and rearrange at the surface. Thus, we can study the influence of the charge density without risks of freezing the conformations. To our knowledge this is the first time that charged lipid membranes have been used to observe supercoiled DNA conformations by AFM. In contrast to mica, a membrane is a soft, elastic substrate for the deposition of DNA. It allows the molecule to rearrange without denaturation. On mica, in the presence of divalent cations, the location of the surface charges is constrained by the crystal lattice. On membranes, charges are fairly mobile.

We study the influence of the surface charge density on the conformation of supercoiled DNA. We observe various 2D equilibrium conformations: plectonemic conformations as well as open conformations or loose conformations with only a few crossings. This only depends on the amount of charges around the molecule, either in volume or at the surface. We show that in a similar manner to salt in bulk, surface charges screen the charge on the DNA backbone resulting in an evolution of supercoiled DNA conformations to more regular and more writhed when surface charge is increased. We compare the efficiency of surface screening and bulk screening (from salt), using two sets of experiments.

## MATERIALS AND METHODS

### AFM

All experiments were performed in tapping mode in the liquid cell of a multimode AFM connected to a nanoscope III controller (Bruker, Santa Barbara). The cantilevers were TAP75 from "Nano and more". To image DNA, a very gentle force was applied. By using small attenuations of the working amplitude we prevent deformation or degradation of the molecule by the tip.

### Substrate

Because of its atomic smoothness, mica was chosen for the bare substrate. However, it is negatively charged, therefore it needs to be modified to promote DNA adhesion. To bridge the macromolecule and the mica, we have chosen charged supported lipid bilayers as a substrate. These model membranes are formed by vesicle fusion on mica (44). By adjusting the ratio between a positively charge lipid and a neutral phospholipid, and assuming that the proportion of charges within the bilayer is the same as within the vesicle solution, we can vary the surface charge. Liposomes of diameter 30 nm on average, as determined by dynamic light scattering, are formed by tip sonication of a solution of the desired lipids mixture diluted in water (0.24 mg/mL) or buffer with the same salt concentration used while imaging. The two lipids chosen are Di-Palmytoyl-Trimethyl-Amonium-Propane (DPTAP), and Di-Palmtoyl-Phosphatidyl-Choline (DPPC). They only differ by their polar head, and no phase separation is expected within the bilayer. Their fluid and gel transitions are respectively 41 and 46°C, therefore we heated up the vesicle solution during sonication, slightly above the fluid gel transition. The hot solution was injected immediately after sonication into the fluid cell of the AFM. When vesicles break on the cold mica, they form instantaneously a bilayer in the gel phase. The membrane was rinsed with 3 mL water or buffer. The obtained supported bilayers are particularly smooth, even smoother than the bare substrate because they partially absorb the corrugation of the mica. However, due to the mechanism of formation of the bilayer, i.e. small domains nucleating and growing, frontiers of domains and sometimes small holes showing the bare mica are clearly visible on the AFM images.

### DNA

The DNA used for this study is supercoiled pUc19 2686 base pairs from New England Biolabs (NEB). It was aliquoted as received (1mg/mL in10mM Tris buffer), without any further purification. This supercoiled DNA appears to be particularly pure of broken or relaxed circles. The manufacturer gives a specification of at least 90% of the DNA in the supercoiled form but agarose gel electrophoresis (1%) performed on several batches in presence of ethidium bromide showed only one bright spot indicating that apparently no relaxed, linear DNA or other sizes of supercoiled DNA were present. 1 μg was unfrozen a few minutes before each experiment and diluted in ultra-pure milliQ water (resistivity 18.2 MΩ.cm) or in buffer solutions to reach a final concentration of 1μg/mL. The DNA dilution was injected into the AFM fluid cell (400μL) and allowed to adsorb gently on the membrane. Imaging of the conformation of the DNA was performed immediately after. To assess the exact linking number of our DNA, we performed 1D agarose gel (1.5 % w/w) electrophoresis in presence of chloroquine (2μg/mL) at 20V for 17h. Chloroquine is an intercalating agent inducing positive superturns and thus reducing negative supercoiling of DNA. An appropriate concentration allows a separation of the topoisomers (Fig. 1).

As a control, we injected at the same time relaxed pUc19 (one strand nicked with NtbspqI from NEB) and relaxed pUc19 that was afterward religated with T4 DNA ligase. Both restriction and ligation were done following the manufacturer protocol. The DNA was purified using a Miniprep PCR purification kit (Qiagen) and resuspended in 10 mM Tris buffer at pH 7.5 before being run through the gel. The conformation of relaxed supercoiled DNA (one strand nicked) is not modified by the presence of chloroquine. As it is circular, its hydrodynamic radius is large and the migration on the gel is the slowest (lane 2). Relaxed DNA that has been religated is positively supercoiled and migrates the fastest (lane 1). The spot of the native pUc19 appears in between (lane 3). A line profile of lane3 is given on figure 1.B. Fitting the intensities with a skewed Gaussian curve gave on averaged, $\Delta L_k = -7$. This value is in agreement with the literature (36).

## RESULTS AND DISCUSSION

### Deposition of supercoiled DNA from pure water on charged membranes

In a first set of experiments we have studied the conformation of supercoiled pUc19 that we allow to adsorb on lipid bilayers with surfaces charges ranging from 0% to 100%. As for surface charges, we refer to the DPTAP content of the bilayer which is the percentage of charged lipids. For the purpose of the experiment we get away from physiological conditions and dissolve our DNA in milliQ water. Working in salt free solution allows us to explore a much larger range of DNA-surface interactions. It is commonly assumed that DNA is not stable in pure water at room temperature (45) (46). But the kinetics of denaturation is still an issue. We used the hypochromism of DNA to determine the kinetics of denaturation of pUc19 DNA in the conditions of our AFM experiments (47). The absorbance of the solution at 260 nm was stable for more than twelve hours at RT. It increased rapidly and reached a plateau when heating the sample up to 80°C (See supplementary information fig. S1). This result clearly shows that the plasmid we used is stable enough at RT for the time of our AFM experiment but that it denatures very fast at 80°C. The fact that plasmids are supercoiled might also prevent a complete denaturation at large scale.

Typical conformations of supercoiled pUc 19 DNA in milliQ water are presented on Fig. 2. For each surface charge, we present a zoom on one single molecule, but all conformations within the population studied look relatively the same, with of course some variations.

On DPPC (fig 2A), whose surface is zwitterionic, plasmids adopt an open circular conformation. Plasmids bearing a negative charge and approaching a zwitterionic membrane, which has a very high dielectric contrast with water, should be repelled by their own image charge. As a consequence there must be some kind of compensating short range interaction that allows for adhesion. The origin of this short range attraction could come from the fact that the lipid dipole is pointing out of the membrane. In such a geometry pUc19 DNA with a $\Delta L_k$ of -7 is particularly constrained. Nicked circles also adopt the same open circular conformation on DPPC. However the circles are not nicked as confirmed by electrophoresis analysis. The same experiment performed with salt in solution gives rise to obvious writhing of the plasmid.

On DPPC-DPTAP mix bilayers, as seen on Fig 2B to 2E, writhing is the consequence of the surface charge evolution of the surface. For low surface charges, the plasmids adopt a relatively loose conformation, which tightens along with an increase of the surface charge density. The superhelix writhes more and more until it reaches the plectonemic conformation that was so often reported in the literature. In contrast relaxed pUc19 DNA stays perfectly open on charged surfaces as one can see on Fig. 3A.

All surface interactions are likely to be similar for the nicked circular DNA and for the native plasmid. Hence for closed circular DNA, the additional internal stress due to extra turns is responsible for writhing at high charge densities. From 10% up to 50% of surface charges, we observe that crossings are right-angled which is a direct evidence of that electrostatic interaction is the most important energy cost for the molecule. Indeed, in this configuration, the chain has to bend a lot to change direction. On the 100 % charges DPTAP surface, we see that the molecule favors the plectonemic conformation with an angle between strands of 30° at crossings. This shows that bending has now become the most expensive energy cost.

Next to the AFM images on Fig. 2, we present the statistical analysis of the mean number of crossings of one strand above or below the other (nodes). This number of nodes increases with the surface charge. For 100% surface charge, it was rather difficult to count the crossings as the superhelix was very tight at the surface and the AFM tip had trouble resolving the tertiary structure. Moreover, for this surface, the error bar is quite large due to surface inhomogeneities. On highly charged surfaces, we observed the appearance of lipid domains in the membrane. There are domains of two different heights. In the thinner domains, lipids are tilted. Their size does not change over time and they also exist at 100% of DPTAP, so we think it is not a demixion transition, but a consequence of the bilayer formation mechanism. Discarding this additional surface complexity we averaged the number of nodes over the entire surface.

In 2D, plasmids are flat almost everywhere except at crossings. Writhe is only due to crossings and each of them contributes to one unit in the writhe number. Each crossing takes away one turn of twist.

Let us assume that in salt free solution for constrained polyelectrolytes adsorbed in 2D, the electrostatic interaction at crossing is dominant compared to other interactions.

On DPPC, the zwitterionic surface, the internal electrostatic repulsion is extremely high. The electrostatic energy for crossing two strands is so expensive, that it overcomes bending and twisting costs. Hence, a DNA molecule prefers to be under high twist rather than supercoiled. Rings are more favorable than "figures of eight", even though the DNA is highly constrained. (See supplementary information for an estimation of the free energy costs with simple arguments and ref 50 for refined calculations).

Experimentally however, the occurrence of the "figure of eight" was not null. This might be due to bridging of strands by nano liposomes not washed out. It may also be crossings that were present in 3D but that the chain did not manage to undo.

One can explain the mechanism of writhing along with a surface charge increase as follows: When charges are incorporated in the bilayer, they screen DNA charges and thus reduce the electrostatic repulsion at crossings. We have noticed that nodes shift along the chain, presumably to reach the most stable conformation in 2D. However, it appears impossible to remove crossings once they have formed. To remove a crossing, desorption of the whole loop is required, which costs too much of energy. We assume that in bulk pure water as there is no salt in solution, DNA should be in a quite open or loose conformation. When in the vicinity of a charged surface, it feels the surface potential and start writhing. This must be done in 3D. When supercoiled DNA lays on the surface, it may still try to decrease the energy costs, but it has not as much freedom to do so. Usually, we observe that nodes move until the DNA form small loops. At equilibrium, supercoiled DNA molecules end up in conformations with small loops scattered along a main cycle.

On highly-charged surfaces, when electrostatic interactions are fully screened, introducing an extra superhelical turn is more favorable than having torsion within the chain. Crossing has become less expensive than bending. In our experiments, it is clear that in that case, plectonemic conformations are observed.

More precise consideration of electrostatics effects shows indeed that the surface charges provide some screening and play a similar role to ionic charges in solution. At the surface carrying $\mu$ charges per unit area, DNA chains adopt a flat conformation and are confined within a counter-ion layer whose thickness is characterized by the so-called Gouy-Chapman (GC) length $\lambda$ from the surface. Within a mean field approach, and in salt free solution, the strength of the surface potential is constant over $\lambda$ and for larger distances $z$ from the surface it decreases as $1/(z+\lambda)$ (48,49). The effective charge of the plasmid is fixed by the interplay between electrostatic interactions and entropy. In solution, cations condense on the backbone of the plasmids. This counter-ion layer reduces significantly the repulsive electrostatic interaction between charged phosphate groups of DNA chains.

For DPPC, close to the surface, there are some cations around the DNA. Their condensation is enhanced by the image charge of the plasmid, which is negative as well but the underlying mechanism is complex.

For charged surfaces, the surface charge reduces the electrostatic interaction between DNA charges. The electrostatic potential of the surface repels cations condensed on the plasmid. They will be released to a large extent. There is an additional, somewhat stronger, screening by the annealed surface charges. Mobile surface charges accumulate under the DNA (Fig. 3.B). As a result, the net charge of the plasmid is higher than for a plasmid subjected to counter-ion condensation in solution. (50,51).

Surface charges shall play a similar role to salt ions in solution and screen the long range electrostatics interactions.

**Comparison of the conformation with charges at the surface or in volume**

To convince ourselves, one may estimate with a simple model an effective charge density to compute an effective screening length in the case of surface screening. This is a preliminary rough estimation that does not take into account the fact that charges at the surface are mobile. We will refine the model later on.

The potentials to take into account are the Debye-Huckel potential (DH) for bulk screening and the GC potentials for surface screening; the corresponding screening lengths being respectively the Debye length $\kappa^{-1}$ and the GC length $\lambda$.

The GC length is given by : $\lambda = \dfrac{1}{2\pi \cdot l_B \cdot \mu}$ with $l_B$ the Bjerrum length (7Å in pure water) .

$l_B = \dfrac{e^2}{4\pi\varepsilon_0\varepsilon_r k_b T}$ with $\varepsilon_r$ the relative dielectric constant of the medium, $\varepsilon_0$ is the vaccum permittivity and $e$ is the elementary charge.

The definition of the Debye length takes into account charges from ions (co ions and counter ions) as well as surface charges.

$\kappa^2 = 4\pi l_B (C_{lipid} + C^+ + C^-)$ where $C_{lipid}$ is the mean positive charge density per unit of volume coming from the surface, and $C^+$ and $C^-$ are respectively the cations and the anions charge density within the GC layer. $C_{lipid}$ is given by $\dfrac{\mu}{\lambda}$ where $\mu$ has been defined before as the surface charge density.

In the GC theory, only one half of the total counter ions is present inside the GC layer (49). Their concentration is then $C^- = \dfrac{\mu}{2\lambda}$. The concentration of the co-ions is zero within the GC layer (except those condensed on the DNA backbone). $C^+ = 0$

$\kappa^2 = 4\pi l_B (C_{lipid} + C^-) = 4\pi l_B \dfrac{3\mu}{2\lambda}$, we obtain $\kappa^{-1} = \dfrac{\lambda}{\sqrt{3}} \approx 0.58\lambda$. So we retain $\kappa^{-1} \propto \lambda$

To compare the effect of surface charges with the effect of charges in bulk, we have performed a second set of experiments on the zwitterionic DPPC surface with addition of salt in solution. This set of experiments showed, as expected, supercoiled conformations with an increase of the number of nodes when the salt concentration was increased. We present in accompanying histograms the quantitative analysis of the node numbers (Fig. 5). As already shown by others in the literature, the conformation of supercoiled DNA shows more writhe as the salt concentration of the solution is increased. One can also notice that the conformations we observe here are slightly different from those screened by the surface. Adhesion to the substrate was rather weak above 10 mM of salt and we encountered a lot of trouble imaging plasmids at 30mM. Plasmids appeared to be whitish on the image presented here, showing that they were almost not touching the substrate. However, the complete scale being 0-5 nm, the maximal height of the plectonemes did not exceed 3 nm.

From the two sets of experiments we have performed we can compare the effect of screening from charges at the surface and from charges in bulk.

By plotting the number of nodes $n$, which is an indicator of the screening efficiency, as a function of the screening length, we expect to see the same behavior for surface screening and bulk screening. This is presented in Fig. 4.B. The area per lipid head was assumed to be 60 $\mathring{A}^2$. As for surface screening, we present only the data points in a moderate range of screening, i.e. where we can rely on a mean field approach. The transition between strong coupling and weak coupling is between 15 and 30 % of charges, but as it is not a strict limit, we plotted the value at 30 % as well ($\lambda$=4.5$\mathring{A}$) as it helps to see a trend.

A minimization of the total free energy of the plasmids with respect to $n$ shows that the free energy has a linear dependence on $n$ in this regime (50). $-n = cst \cdot (F_h + F_{cross}) - \Delta L_k$ where $F_h$ is the interaction with the surface and $F_{cross}$ is the penalty for crossing helices in 2D. In the mean field regime, $F_{cross}$ is dominant. A linear dependence with $\lambda$ is hidden in $F_{cross}$. We find in good agreement with the theory that the number of nodes decreases linearly with the screening length. We do see that the point at 30% has already started to deviate from the mean field regime. The number of nodes is zero when $F_{cross}$ is too high (for large screening length).

The intercept should give us $-\Delta L_k$. Using the two data points in the mean field regime, we find an intercept at 6.5 which is rather close to the value of $\Delta L_k$ =-7 found by gel electrophoresis. For bulk screening, we see a similar behavior. $n$ is also decreasing but less sharply. As it was impossible to image plasmids above 30 mM of salt, we have only a couple of data points to compare with surface screening. We added an extra point at zero screening length, i.e. when the electrostatics is completely screened. This data point should be common to both types of screening if one only considers the electrostatic cost. This point at zero screening length is only an extrapolation of the mean field regime and is not an experimental measure. In reality, for highly charged surfaces, other types of interactions do play a role (for instance bending) and it is not obvious that surface screening and bulk screening data should merge.

The straight line based on the two data points ($\kappa^{-1}$= 0 and $\kappa^{-1}$= 30 $\mathring{A}$) crosses the $x$ axis at $\kappa^{-1}$= 50 $\mathring{A}$. It is then quite surprising that we observed a mean node number of one for $\kappa^{-1}$=96$\mathring{A}$. At this distance the electrostatic cost for crossing helices in 2D is of the order of 100 $k_B$T (19) (52). We thus believe that this crossing is the result of frozen 3D projection. In bulk, a crossing is still possible as soon as the strands are sufficiently parted (at least by one Debye length). In contrast to surface screening where plasmids have to coil when they feel the surface potential, here plasmids have to uncoil when they already are at the surface. Even in the absence of salt, at $\kappa^{-1}$=960 $\mathring{A}$, with a penalty of crossing of the order of 1000 $k_B$T we also observed some "figure of eight" molecules. Their number being less than 20% of the total, their presence did not strongly affect $n$. It seems to be more difficult to uncoil to avoid a crossing than coiling to make one. Therefore it is likely that we have in our bulk statistics crossings that are 3D projections. It is reasonable to think that it affects only small fractions of nodes and that for sufficiently high numbers, it is probably negligible. To compare bulk and surface screening we must discard the point at $\kappa^{-1}$=96 Å. By comparing the slope of the two adjusted straight lines, one obtains the equivalence $\frac{S_{\kappa^{-1}}}{S_\lambda} = 2.36$, which is much larger than the factor $\frac{1}{\sqrt{3}}$ we found in our first theoretical estimation.

This discrepancy has several origins. First, the DH and GC screening lengths have qualitatively the same effect, but the two corresponding fields do not have the same dependence on the distance. Second we did not take into account the surface charges annealing. In consequence, one needs to make a more quantitative comparison. As we mentioned before, the most important interaction is the interaction at crossings. We compute the interaction potential between two crossing segments, in the case of bulk screening and surface screening.

The free energy can be written as a function of a 2D Fourier Transform (FT) of the potential divided by the $sin$ of the angle between segments.

$$F_{cross} = \frac{\tilde{V}(q=0, z=0)}{\sin\theta}$$

The minimum of electrostatic free energy is obtained when the segments are perpendicular to each other as it is observed at moderate surface charge. We consider this case in the following discussion.

To get the correspondence between the two screening distances, we need to compare the energy cost per crossing.

For 3D, The DH equation applies with a source term corresponding to one charge. We expect the potential to be of Yukawa type.

$$V_{DH}(r) = \frac{l_B}{r} e^{-\kappa r} \rho_\kappa^2$$ with $r$ the radial distance from the charge and $\rho_\kappa$ the linear charge density of the double helix. In order to compare the free energies for DH and GC, we need to compute the potentials in momentum space. By applying a FT to the Yukawa potential and writing it at the surface, we easily get $\tilde{V}_{DH}(q=0) = 2\pi l_B \kappa^{-1} \rho_\kappa^2$

The same way, the GC potential should scale with the square of the linear charge density of the DNA. This interaction is not as easy to compute. It has been calculated exactly in ref (50).

$$\tilde{V}_{GC}(q=0) = 4\pi l_B \rho_\lambda^2 \lambda$$ with $\rho_\lambda$ the linear charge density of DNA in the case of surface screening. In the presence of annealed charges due to the mobility of lipids within the bilayer, the GC potential is reduced by a factor of 3.

Hence, the energy penalties for crossing are respectively $4\pi l_B \overline{\rho_\kappa}^2 \lambda$ , $\frac{4\pi l_B \overline{\rho_\lambda}^2 \lambda}{3}$ , $2\pi l_B \rho_\kappa^2 \kappa^{-1}$ for quenched surface charges, annealed surface charges and neutral surfaces in the presence of salt:

Coming back to Fig.5 showing the node number as a function of the screening lengths and comparing the slope $S_{\kappa^{-1}}$ (for bulk screening) and $S_\lambda$ (for surface screening) of the two straight lines, we found $S_{\kappa^{-1}} = 2.36 S_\lambda$.

If the two DNA segments are screened the same way by a DH or by a GC potential, the value of the two interactions are equal. This suggests that the ratio of effective charges is $\frac{2}{3}\rho_\lambda{}^2 \Big/ \rho_\kappa{}^2 = 2.36$ and then $\frac{\rho_\lambda}{\rho_\kappa} \approx 1.9$. The linear charge density of DNA appears to be two times smaller for bulk screening than for surface screening, which is compatible with our assumption that DNA has less condensed couterions on charged surfaces than in bulk.

The point that we have not considered yet is the difference of adsorption mechanisms on the two surfaces, the zwitterionic one and the charged one.

On the neutral bilayer (zwitterionic), counterion condensation is slightly enhanced while on the charged bilayer, there is counterion release. Indeed on the zwitterionic bilayer, to compensate for the repulsion of plasmids by their own image charge, there is overcondensation of counterions, which reduces the effective charge of DNA a little further.

A more precise analysis would lead to the real proportion of counterions effectively condensed on the DNA backbone for the two cases. As we have only a couple of valid data points in the mean field regime and many sources of uncertainty, we did not carry on the analysis further.

We have mainly focused our study on electrostatics interactions and did not point out the question of the entropy. In the experiments that we have shown we are either in salt free solution or in the moderate salt concentration regime. When electrostatic repulsion is high, entropy is certainly negligible, which is mostly the case in our experiments, however, in physiological conditions, entropy may be important. Nonetheless it should not alter the node number $n$.

In order to conclude on the 2D entropic effect on the conformation of supercoiled DNA, for instance on more or less writhe as suggested by Velichko's work, we would need an experiment that emphasizes the pure effect of the entropy. The easiest way would be to compare 2D conformations in high salt concentration where the electrostatic repulsion is screened, with 3D bulk conformations. Unfortunately, this is not possible. Up to now, no experimental technique is available to give a real photography of the conformation of supercoiled DNA in 3D.

The only obvious influence of the entropy we have observed in our experiments is the occurrence of very small loops next to much larger ones. Indeed, if the crossings are well scattered all over the DNA molecule just after immobilization, their positions evolve during the first minutes. We clearly see that after some time, there is equilibrium of very small loops next to much larger ones. This effect has been previously reported with computer simulations on 2D knots (53,54).

**CONCLUSION:**

In this article we have presented experiments realized on pUc19 plasmid DNA adsorbed on a bilayer and changed the surface charge density to see its effect on the 2D supercoiled conformation of the DNA. We have been able to reproduce many conformations reported in literature either by AFM or by TEM, just by changing the surface charge. Supercoiled DNA might be completely open as a ring in the absence of charges at the surface and in bulk. It coils when we increase the amount of charges at the surface to finally reach the plectonemic superhelix conformation which is an analogue of the tertiary structure often reported in TEM experiments. The most important energetic cost is the repulsion at crossing which cannot be avoided in 2D. In the absence of screening charges, in bulk or at the surface, the molecule favours open circular conformations, even though it remains very constrained by the linking number. When charges are added at the surface, the GC potential screens charges on the DNA backbone, and supercoiling occurs. The presence of mobile charges at the surface does play a very important role. It first ensures the 2D equilibrium of the molecule but it also enhances screening at the surface. By comparing the effect of surface screening and bulk screening for a molecule adsorbed at the surface, we extracted the ratio of linear charge density of the DNA between surface screening and bulk screening which is the direct consequence of different ionic condensation mechanisms.

Membranes of other lipid compositions can also be a good substrate to study the dynamics of macromolecules or macromolecules at equilibrium (55). The interaction between charged lipids and biomolecules of opposite charges is also of interest. In the present case, the DNA is negatively charged and the membrane underneath is positively charged. In protein- membrane interactions, both charges are of reversed sign. Cell membranes are negative and objects that bind to them are positive. Nevertheless, pure electrostatic effects can lead to similar consequences.

## SUPPORTING MATERIAL
Are available at

## ACKNOWLEDGMENTS
The authors thank Albert Johner for fruitful discussions and critical corrections of the manuscript.

## REFERENCES:

1.  Dulbecco, R. and M. Vogt. 1963. Evidence for A RIing Structure of Polyoma Virus DNA. Proceedings of the National Academy of Sciences 50:236-243.

2.  Cozzarelli, N. R. 1980. DNA gyrase and the supercoiling of DNA. Science 207:953-960.

3.  Wasserman, S. A. and N. R. Cozzarelli. 1986. Biochemical topology: applications to DNA recombination and replication. Science 232:951-960.

4.  Kornberg, A. 1988. DNA replication. Journal of Biological Chemistry 263:1-4.

5.  Postow, L., N. J. Crisona, B. J. Peter, C. D. Hardy, and N. R. Cozzarelli. 2001. Topological challenges to DNA replication: Conformations at the fork. Proceedings of the National Academy of Sciences of the United States of America 98:8219-8226.

6.  Gellert, M. 1981. DNA Topoisomerases. Annual Review of Biochemistry 50:879-910.

7.  Horowitz, D. S. and J. C. Wang. 1984. Torsional rigidity of DNA and length dependence of the free energy of DNA supercoiling. Journal of Molecular Biology 173:75.

8.  Witz, G. and A. Stasiak. 2010. DNA supercoiling and its role in DNA decatenation and unknotting. Nucleic Acids Research 38:2119-2133.

9.   Crick, F. H. 1976. Linking numbers and nucleosomes. Proceedings of the National Academy of Sciences 73:2639-2643.

10.  Boles, T. C., J. H. White, and N. R. Cozzarelli. 1990. Structure of plectonemically supercoiled DNA. Journal of Molecular Biology 213:931.

11.  Wang, J. C. 1974. Interactions between twisted DNAs and enzymes: The effects of superhelical turns. Journal of Molecular Biology 87:797.

12.  Depew, D. E. and J. C. Wang. 1975. Conformational fluctuations of DNA helix. Proceedings of the National Academy of Sciences 72:4275-4279.

13.  Anderson, P. and W. Bauer. 1978. Supercoiling in closed circular DNA: dependence upon ion type and concentration. Biochemistry 17:594.

14.  Adrian, M., B. ten Heggeler-Bordier, W. Wahli, A. Z. Stasiak, a. Stasiak, and J. Dubochet. 1990. Direct visualization of supercoiled DNA molecules in solution. The EMBO journal 9:4551.

15.  Sperrazza, J. M., J. C. Register Iii, and J. Griffith. 1984. Electron microscopy can be used to measure DNA supertwisting. Gene 31:17.

16.  Brack, C. 1981. DNA electron microscopy. CRC critical reviews in biochemestry:113.

17.  Bednar, J., P. Furrer, A. Stasiak, J. Dubochet, E. H. Egelman, and A. D. Bates. 1994. The Twist, Writhe and Overall Shape of Supercoiled DNA Change During Counterion-induced Transition from a Loosely to a Tightly Interwound Superhelix: Possible Implications for DNA Structure in Vivo. Journal of Molecular Biology 235:825.

18.  Vologodskii, A. V. and N. R. Cozzarelli. 1994. Conformational and Thermodynamic Properties of Supercoiled DNA. Annual Review of Biophysics and Biomolecular Structure 23:609-643.

19.  Schlick, T., B. Li, and W. K. Olson. 1994. The influence of salt on the structure and energetics of supercoiled DNA. Biophysical journal 67:2146.

20.  Langowski, J., U. Kapp, K. Klenin, and A. Vologodskii. 1994. Solution structure and dynamics of DNA topoisomers: Dynamic light scattering studies and Monte Carlo simulations. Biopolymers 34:639.

21.  Gebe, J. A. and J. M. Schurr. 1996. Thermodynamics of the first transition in writhe of a small circular DNA by Monte Carlo simulation. Biopolymers 38:493.

22.  Rybenkov, V. V., A. V. Vologodskii, and N. R. Cozzarelli. 1997. The effect of ionic conditions on the conformations of supercoiled DNA. I. sedimentation analysis. Journal of Molecular Biology 267:299.

23. Rybenkov, V. V., A. V. Vologodskii, and N. R. Cozzarelli. 1997. The effect of ionic conditions on the conformations of supercoiled DNA. II. equilibrium catenation. Journal of Molecular Biology 267:312.

24. Mehmet, S. and et al. 2010. Twist-writhe partitioning in a coarse-grained DNA minicircle model. Physical Review E 81:041916.

25. Bustamante, C., J. Vesenka, C. L. Tang, W. Rees, M. Guthold, and R. Keller. 1992. Circular DNA molecules imaged in air by scanning force microscopy. Biochemistry 31:22.

26. Schaper, A., J. P. Starink, and T. M. Jovin. 1994. The scanning force microscopy of DNA in air and in n-propanol using new spreading agents. FEBS Lett 355:91.

27. Samori, B., G. Siligardi, C. Quagliariello, A. L. Weisenhorn, J. Vesenka, and C. J. Bustamante. 1993. Chirality of DNA supercoiling assigned by scanning force microscopy. Proceedings of the National Academy of Sciences 90:3598-3601.

28. Bezanilla, M., S. Manne, D. E. Laney, Y. L. Lyubchenko, and H. G. Hansma. 1995. Adsorption of DNA to Mica, Silylated Mica, and Minerals: Characterization by Atomic Force Microscopy. Langmuir 11:655.

29. Lyubchenko, Y. L. and L. S. Shlyakhtenko. 1997. Visualization of supercoiled DNA with atomic force microscopy in situ. Proceedings of the National Academy of Sciences 94:496-501.

30. Rippe, K., N. Mücke, and J. r. Langowski. 1997. Superhelix dimensions of a 1868 base pair plasmid determined by scanning force microscopy in air and in aqueous solution. Nucleic Acids Research 25:1736-1744.

31. Tanigawa, M. and T. Okada. 1998. Atomic force microscopy of supercoiled DNA structure on mica. Analytica Chimica Acta 365:19.

32. Zuccheri, G., R. T. Dame, M. Aquila, I. Muzzalupo, and B. Samorì. 1998. Conformational fluctuations of supercoiled DNA molecules observed in real time with a scanning force microscope. Applied Physics A: Materials Science & Processing 66:S585.

33. Bussiek, M., N. Mucke, and J. Langowski. 2003. Polylysine coated mica can be used to observe systematic changes in the supercoiled DNA conformation by scanning force microscopy in solution. Nucleic Acids Research 31:e137.

34. Fogg, J., M. and et al. 2006. Exploring writhe in supercoiled minicircle DNA. Journal of Physics: Condensed Matter 18:S145.

35. Billingsley, D. J., J. Kirkham, W. A. Bonass, and N. H. Thomson. 2010. Atomic force microscopy of DNA at high humidity: irreversible conformational switching of supercoiled molecules. Physical Chemistry Chemical Physics 12:14727.

36.   Zakharova, S. S., W. Jesse, C. Backendorf, S. U. Egelhaaf, A. Lapp, and J. R. C. van der Maarel. 2002. Dimensions of Plectonemically Supercoiled DNA. Biophysical journal 83:1106.

37.   Dubochet, J., M. Adrian, J.-J. Chang, J.-C. Homo, J. Lepault, A. W. McDowall, and P. Schultz. 1988. Cryo-electron microscopy of vitrified specimens. Quarterly Reviews of Biophysics 21:129-228.

38.   Gebe, J. A., J. J. Delrow, P. J. Heath, B. S. Fujimoto, D. W. Stewart, and J. M. Schurr. 1996. Effects of Na+and Mg2+on the Structures of Supercoiled DNAs: Comparison of Simulations with Experiments. Journal of Molecular Biology 262:105.

39.   Vologodskii, A. V., S. D. Levene, K. V. Klenin, M. Frank-Kamenetskii, and N. R. Cozzarelli. 1992. Conformational and thermodynamic properties of supercoiled DNA. Journal of Molecular Biology 227:1224.

40.   Velichko, Y. S., K. Yoshikawa, and A. R. Khokhlov. 2000. Surface-Induced DNA Superhelicity. Biomacromolecules 1:459.

41.   Fujimoto, B. S. and J. M. Schurr. 2002. Monte Carlo Simulations of Supercoiled DNAs Confined to a Plane. Biophysical Journal 82:944.

42.   Rivetti, C., M. Guthold, and C. Bustamante. 1996. Scanning Force Microscopy of DNA Deposited onto Mica: EquilibrationversusKinetic Trapping Studied by Statistical Polymer Chain Analysis. Journal of Molecular Biology 264:919.

43.   Scomparin, C., S. Lecuyer, M. Ferreira, T. Charitat, and B. Tinland. 2009. Diffusion in supported lipid bilayers: Influence of substrate and preparation technique on the internal dynamics. Eur. Phys. J. E 28:211-220.

44.   Richter, R., A. Mukhopadhyay, and A. Brisson. 2003. Pathways of Lipid Vesicle Deposition on Solid Surfaces: A Combined QCM-D and AFM Study. Biophysical journal 85:3035.

45.   Owen, R. J., L. R. Hill, and S. P. Lapage. 1969. Determination of DNA base compositions from melting profiles in dilute buffers. Biopolymers 7:503.

46.   Frank-Kamenetskii, M. D. 1971. Simplification of the empirical relationship between melting temperature of DNA, its GC content and concentration of sodium ions in solution. Biopolymers 10:2623.

47.   Voet, D., W. B. Gratzer, R. A. Cox, and P. Doty. 1963. Absorption spectra of nucleotides, polynucleotides, and nucleic acids in the far ultraviolet. Biopolymers 1:193.

48.   Adamson, A. W. 1990. Physical Chemistry of Surfaces. New York: John Wiley and Sons Inc.

49.  Andelman, D. 1995. Electrostatic Properties of Membranes: The Poisson-Boltzmann theory. In Structure and Dynamics of Membranes. R. Lipowsky and E. Sackmann, editors. Elsevier Science B.V. North Holland.

50.  Lee, N. K., T. Schmatko, P. Muller, M. Maaloum, and A. Johner. preprint. The Shape of adsorbed supercoiled plasmids: An equilibrium description. Physical Review E.

51.  Fleck, C. C. and R. R. Netz. 2007. Surfaces with quenched and annealed disordered charge distributions. The European Physical Journal E: Soft Matter and Biological Physics 22:261.

52.  Randall, G. L., B. M. Pettitt, G. R. Buck, and E. L. Zechiedrich. 2006. Electrostatics of DNA-DNA juxtapositions: consequences for type II topoisomerase function. Journal of Physics: Condensed Matter 18:S173.

53.  Metzler, R., A. Hanke, P. G. Dommersnes, Y. Kantor, and M. Kardar. 2002. Equilibrium Shapes of Flat Knots. Physical Review Letters 88:188101.

54.  Kardar, M. 2008. The elusiveness of polymer knots. Eur. Phys. J. B 64:519-523.

55.  Heimburg, T., B. Angerstein, and D. Marsh. 1999. Binding of Peripheral Proteins to Mixed Lipid Membranes: Effect of Lipid Demixing upon Binding. Biophysical Journal 76:2575.

56.  Baumann, C. G., S. B. Smith, V. A. Bloomfield, and C. Bustamante. 1997. Ionic effects on the elasticity of single DNA molecules. Proceedings of the National Academy of Sciences 94:6185-6190.

## FIGURES TITLES AND LEGENDS

**Fig. 1 : Agarose gel electrophoresis (1%) of pUc 19 DNA in presence of chloroquine (2µg/mL). On left, lane 1: Closed circular pUc19 DNA (Relaxed by nicking and re ligated). Lane 2: open circular (one strand nicked) pUc19 DNA (relaxed by nicking). Lane 3: native supercoiled puc 19 DNA (initially negatively supercoiled). On right, intensity line profile of lane 3.**

**Fig. 2 : A to E show 400 \*400 nm zooms of representative puc19 DNA conformations. Plasmids are diluted in milli Q water and adsorbed on bilayers with a controlled surface charge and containing a proportion of cationic lipid of 0, 10, 30, 50, and 100 %  respectively. F to J : Corresponding histograms of the number of nodes (number of crossings).**

**Fig. 3 : A: Open circular (one strand nicked) pUc 19 DNA adsorbed on a 50% charged membrane. The DNA is lying open on the surface. There is no apparent crossing of strands in the tertiary structure. The bilayer was not complete. Black areas are holes that let appear the mica substrate underneath.**
**B:  Sketch of a charged rod adsorbed on an oppositely charged membrane with mobile charges.**

**Fig.4 : effect of salt addition on a uncharged DPPC bilayer : from A to D, the added salt concentration is respectively 0mM, 1mM, 10mM, 30 mM of NaCl,  from E to G, the corresponding statistical analysis of the number of nodes.  The number of DNA superturns increases along with the bulk salt concentration, but at the same time the affinity for the surface is reduced and imaging becomes very difficult.  The color scale is the same as on Fig.2.**

**Fig. 5 : Comparison of the efficiency of surface screening and bulk screening. We measured the crossing number as a function of the corresponding screening length in the case of the Gouy-Chapman (*red solid circles*) and Debye Huckel (*black solid squares*) theories respectively.**

**FIGURES**

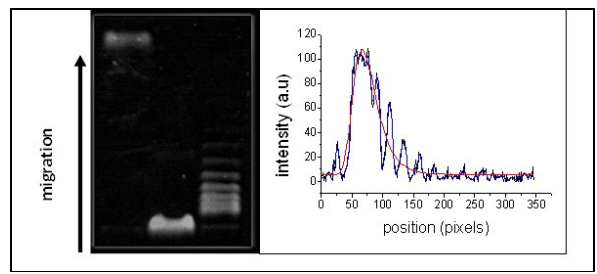

**Fig. 1 :**

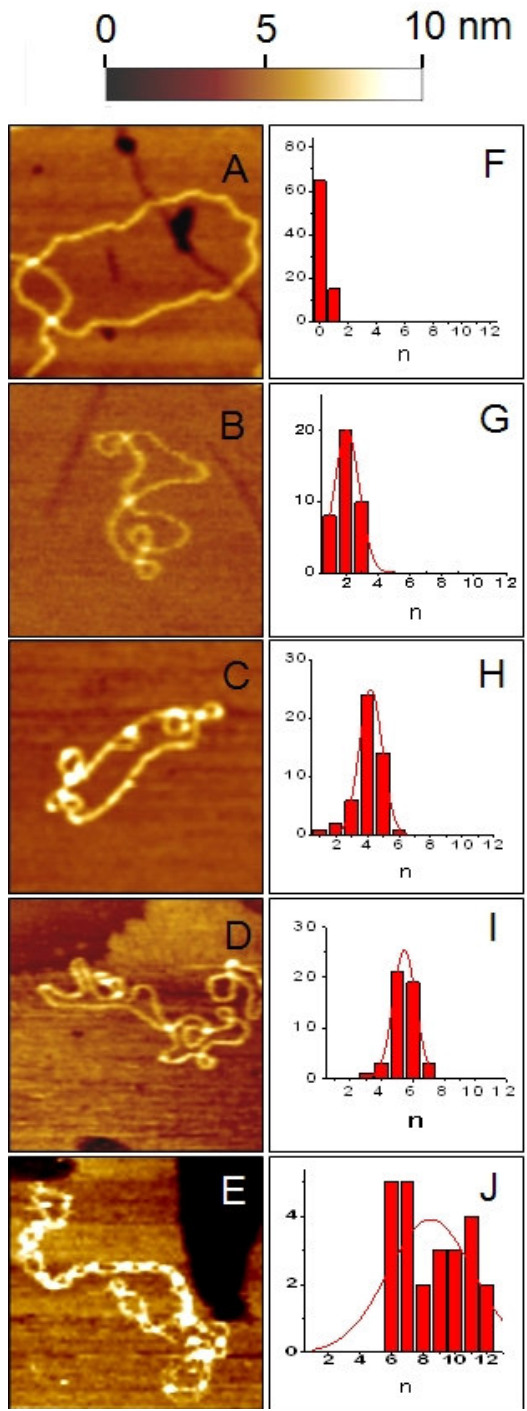

**Fig. 2 :**

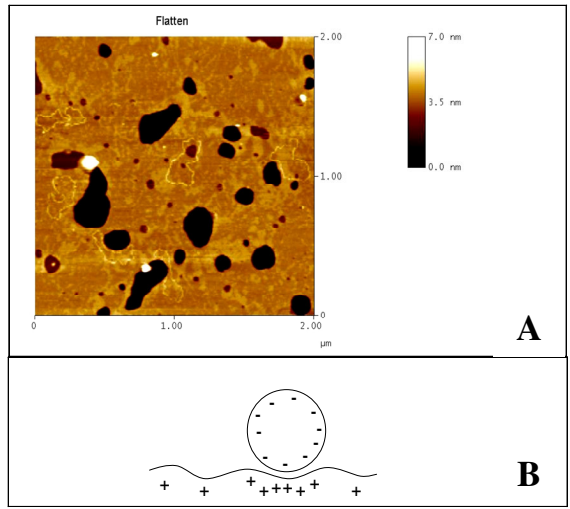

Fig.3 :

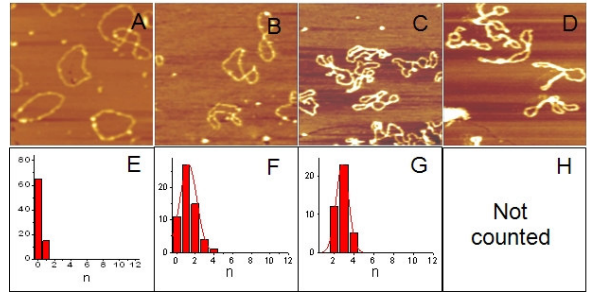

**Fig. 4**

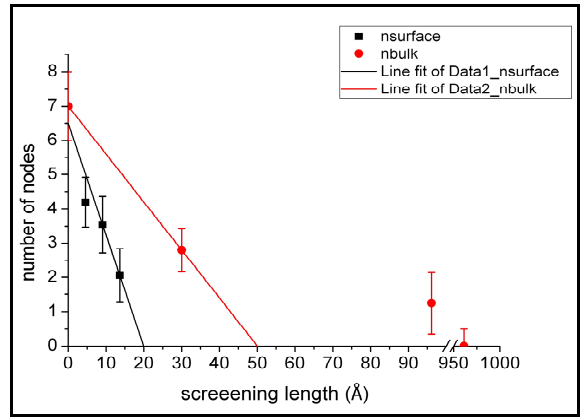

**Fig. 5 :**

**SUPPORTING MATERIAL**

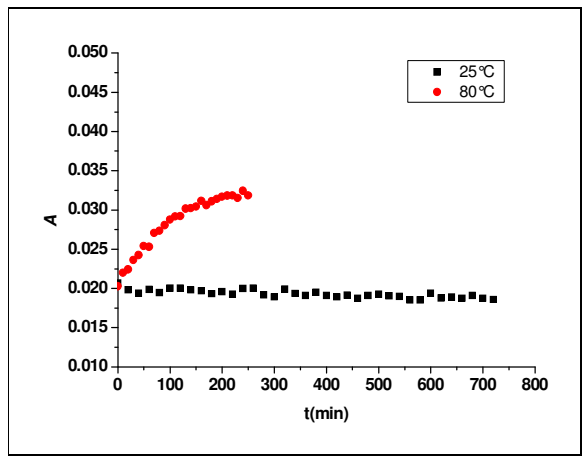

**Fig. S 1: Measurement of the evolution of the absorbance of pUc19 DNA at 260 nm as a function of time.** *(black squares)* **room temperature recording.** *(red circles)* **recording when heating at 80°C.The DNA was dissolved in pure water at a concentration of 1µg/mL which is the concentration used in AFM experiments.**

**S2: Estimations of the free energy costs in 2D and in salt free solution**

To understand the circular form of supercoiled DNA in the absence of charges at the surface and why writhing occurs when charges are added at the surface, we must compare, specifically for 2D, the (free) energetic cost of writhing, twisting, and the electrostatic repulsion in pure water.

As for the entropic cost, it is of the order of a few $k_BT$ and will only become manifest when loops are not too rigid (not too small). The dominating crossing energy is fairly independent of the shape, provided that the crossing angles are optimized.

$$F = F_{elect} + F_{twist} + F_{bending}$$

In the absence of screening, the electrostatic cost of a crossing is overwhelming. To show this, a rough estimate is enough.

To illustrate our explanation, we compare two different conformations that we observed in pure water on DPPC, (in the absence of screening from the surface): a ring of radius R and what we have called a "figure of eight". For simplicity of the calculation, the "figure of eight" is taken as two circles of radius R/2 in close contact but we keep in mind that they do cross strands. All energies are given in $k_BT$ units.

The bending cost is defined as the necessary energy for bending a rod of length L over a circle of radius R. $F_{bending} = \frac{1}{2} l_p \left(\frac{1}{R}\right)^2 L$ with $l_p$ is the persistence length of the DNA.

In water, due to the high repulsion of phosphates on its backbone, DNA should have a greater persistence length than the usual 50 nm in physiological conditions. In the literature, we found 94 nm measured with optical tweezers in 1mM NaCl (56) but nothing for more dilute solutions. In our experiments, we do not have any NaCl but Tris buffer at a very low concentration, roughly $10^{-5}$M. This remaining ionic content comes from the dilution of the stock plasmids solution.

On the DPPC membrane, we have measured in pure water, the persistence length on linear pUc19 DNA obtained by restriction of the supercoiled form. We measured the end to end distance for a hundred molecules in order to compute the average persistence length (42). According to Rivetti et al, if the conformation of the chain is at equilibrium in 2D, the average of the square of the end to end distance is equal to $4 l_p$ L. $\left\langle R^2 \right\rangle = 4 l_p L$

We found experimentally that $l_p = 87$ nm. This is slightly smaller than the 94 nm found in 1mM NaCl and not higher as expected but considering the large error bar the two values are consistent. On this neutral surface, a statistical analysis of the measured contour length gave 860 ±25 nm which is slightly smaller than the theoretical length of 913 nm. This is not anomalous considering the fact that some part of the chain may be out of plane because not well bound to the surface. This effect is very difficult to detect with the fluctuations of the membrane underneath.

For the ring, the bending free energy is $F_{o,bending} = \frac{1}{2} l_p \left(\frac{2\pi}{L}\right)^2 L$ with $L$ the Contour length of the plasmid.

For the figure of eight, the bending free energy is $F_{8,bending} = \frac{1}{2} l_p \left(\frac{4\pi}{L}\right)^2 \frac{L}{2} + \frac{1}{2} l_p \left(\frac{4\pi}{L}\right)^2 \frac{L}{2}$

$$\Delta F_{bending=} 6\pi^2 \frac{l_p}{L} \approx 6 k_B T$$

The free energy of twist is defined as the necessary energy for twisting a rod of length L, of an angle $\alpha$ over a length $l_t$ (the twist length).

$$F_{twist} = \frac{1}{2} l_t \left(\frac{\alpha}{L}\right)^2 L \text{ with } l_t \text{ the twist length and } \alpha \text{ the twist angle. } \alpha = 2\pi T_w$$

In physiological conditions, a value of $l_t = 70$ nm is commonly assumed. To our knowledge there is no data on the variation of $l_t$ with screening (salt concentration). In pure water, we assumed that $l_t$ was not sensitive to screening in contrast to $l_p$.

For the ring, $W_r = 0$, thus $T_w = \Delta L_k$,

$$F_{o,twist} = \frac{1}{2} l_t \left(\frac{2\pi \cdot \Delta L_k}{L}\right)^2 L$$

For the figure of eight, we can consider than the two segments are perpendicular to each other at the crossing and one extra turn contribute for one in $W_r$.

$W_r = -1$ and thus $T_w = \Delta L_k + 1$.

$$F_{8,twist} = \frac{1}{2} l_t \left(\frac{2\pi \cdot (\Delta L_k + 1)}{L}\right)^2 L$$

Using $\Delta L_k = -7$

$$\Delta F_{twist} \approx \frac{1}{2} \frac{l_t}{L} \cdot 2\Delta L_k \cdot 4\pi^2 \approx -17 kT$$

For the electrostatic cost, we define two different contributions to the free energy, the self electrostatic repulsion which is the internal repulsive electrostatic interaction due to the constrained circularity of a plasmid and the electrostatic cost of a perpendicular crossing.

The self electrostatic free energy in $k_B T$ units follows Coulomb's law: $F_{self} = l_B \dfrac{q^2}{R}$ where $l_B$ is the Bjerrum length and is equal to 7 Å in pure water,

$q$ is the total net charge of the plasmid , $q = L \cdot \rho$ with $L$ the contour length and $\rho$ the linear charge of the DNA. We assume one charge every $l_B$.

$R$ is the distance between two interacting segments along the molecule. In solution, two segments may interact if they are separated by less than a Debye length. In our salt situation, it is around 100 nm. Looking at a circle immobilized in 2D, this distance is of the order of the radius of the circle itself. Therefore taking the radius of a big circle (the ring) or the one of two little circles (the figure of eight) will tell us whether there is more energy stored in two little circles than in a big one.

$$F_{o,self} = \frac{l_B}{R_c} \cdot \left(\frac{L}{l_B}\right)^2 = \frac{L \cdot 2\pi}{l_B} \text{ with } R_c = \frac{L}{2\pi}$$

$$F_{8,self} = F_{o+o} = \frac{l_B}{R/2} \left(\frac{L/2}{l_B}\right)^2 + \frac{l_B}{R/2} \left(\frac{L/2}{l_B}\right)^2 = \frac{L \cdot 2\pi}{l_B}$$

The electrostatic self-energy is roughly equal for the two conformations, but we do not have taken into account yet the interaction energy at crossing in the "figure of eight". A rough underestimate is $F_{cross} \approx \dfrac{2l_B}{R/2} \left(\dfrac{1}{l_B} \cdot \dfrac{L}{2}\right)\left(\dfrac{1}{l_B} \cdot \dfrac{L}{2}\right) \approx 2000 k_B T$

This is the electrostatic repulsion of two punctual charges spaced by a distance R and bearing the same total charges than two DNA circles of radius R/2.

In contrast to the bulk case, where writhe may still be possible in loose conformations, in 2D the strands have to be in close contact at crossing. This is a huge energetic cost but we should point out that close contact means much closer than the given Debye length $\kappa^{-1}$ that is 94 nm at $10^{-5}$M of salt. In another article (50) we have estimated the energy of crossing for full screening, which is almost correct with a more refined calculation. $F_{cross} \approx \dfrac{4\pi}{\kappa d_B} \approx 1700 k_B T$

The crossing free energy is relevant in biology for instance when enzymes bring strands into close contact, and has been studied by computation, albeit in salt solution (19,52). Schlick et al data scale with the logarithm of the salt concentration (19). Extrapolating their data to a concentration of $10^{-5}$ M gives a value of $1000 k_B T$ per molecule as well.

Hence, we see that the electrostatic energy for crossing is dominant in salt free solution and that the other contributions to the free energy (writhing, twisting and entropy) are negligible in comparison.